\documentclass[rnote]{aa} % for the research notes

\usepackage{color}
\usepackage{ulem}
\usepackage{graphicx}
\usepackage[varg]{txfonts}
\begin{document}
\title{Parameters of rotating neutron stars with and without hyperons}

\author{M. Bejger}
\institute{N. Copernicus Astronomical Center, Bartycka 18, PL-00-716 
Warsaw, Poland
\\ \email{bejger@camk.edu.pl}}

\date{Received 10/12/2012; accepted 15/02/2013}

\abstract
% context
{The discovery of a $2 M_\odot$ neutron star provided a robust 
constraint for the theory of exotic dense matter, bringing into question the  
existence of strange baryons in the interiors of neutron stars. 
Although many theories fail to reproduce this observational result, 
several equations of state containing hyperons are consistent with it.}
% aims
{We study global properties of stars using equations of state containing 
hyperons, and compare 
them to those without hyperons to find similarities, differences, 
and limits that can be compared with the astrophysical observations.}
% methods
{Rotating, axisymmetric, and stationary stellar configurations in general 
relativity are obtained, and their global parameters are studied.}
% results
{Approximate formul\ae\ describing the behavior of the maximum and minimum stellar mass, 
compactness, surface redshifts, and moments of inertia as functions of spin frequency 
are provided. We also study the thin disk accretion and compare the spin-up evolution 
of stars with different moments of inertia.}
% conclusions
{} 
   
\maketitle
%-----------------------------------------------------------------------------
\section{Introduction}
%-----------------------------------------------------------------------------
The discovery of a $2 M_\odot$ neutron star (NS) \citep{Demorest2010} provides
a strong motivation for (the community of) dense-matter physicists 
to understand the role of exotic phases in the interior of neutron stars.  
The appearance of new
particles (in general, new phases of matter) typically results in the {\it
softening} of matter, since the contribution to pressure from energetic
nucleons is replaced by that of slowly-moving, massive particles; the pressure
increases less steeply with the density in the equation of state (EOS), which in
turn lowers the NS maximum allowable mass (for a review concerning dense matter
EOSs, see \citealt{HaenselPY2007}). The state of the dense
matter in the NS cores is very different from the state known at the energies and
densities reached by the terrestrial experiments; notably, stable strange
matter may appear in the interiors of NSs at densities a few times the nuclear
saturation density, with strange baryons (hyperons) being a very probable result. 

We study a set of selected EOSs, representing current theoretical approaches to
the description of the dense matter, that are consistent with the robust
constraint put forward by the observers. If we assume that hyperons do exist
in the interiors of massive NSs, then based on our present knowledge it seems
necessary for such objects to be composed of sufficiently stiff matter (at
lower densities) to sustain the softening that is introduced by the
appearance of hyperons. In other words, substantial stiffness of the EOS of nucleonic
(non-strange) matter is a necessary condition for the existence of hyperons in
the core of a massive NS. Consequently, stellar configurations that contain
hyperons will, at least partly, differ from those without hyperons. We therefore seek
similarities and differences between them, as well as limits that can be
applied to astrophysical observations. 

This Note is composed as follows: Sect.~\ref{sect:meth_eos} contains 
the description of methods and EOSs used; Sect.~\ref{sect:results} presents 
the results for the gravitational mass, surface redshift, moment of inertia 
and accretion tracks; Sect.~\ref{sect:conc} contains the conclusions and summary.
%-----------------------------------------------------------------------------
%-----------------------------------------------------------------------------
\section{Equations of state}
\label{sect:meth_eos}
%-----------------------------------------------------------------------------
%-----------------------------------------------------------------------------
The following selection of nucleonic EOSs is used: the APR EOS (\citealt{AkmalPR1998},
model $A18+\delta v+UIX^\star$) is a variational, non-relativistic many-body
solution with relativistic corrections; DH EOS \citep{DouchinH2001} is
constructed using a non-relativistic energy density functional based on the
SLy4 effective nuclear interaction, designed to describe both crust and core in 
unified way; the BSK20 EOS (moderate stiffness EOS by \citealt{GorielyCP2010})
is also based on the nuclear energy-density functional theory, using the
generalized Skyrme forces fitted to experimental nuclear data and reproducing
properties of infinite nuclear matter from the calculations of many-body
interactions.

Realistic microscopic dense-matter theories (the Brueckner–Hartree–Fock approach) 
predict maximum masses of NSs with
hyperons much below the observed $\simeq 2 M_\odot$ (see e.g.,
\citealt{SchulzePRV2006}, \citealt{BurgioSL2011}, and references therein). 
This may be caused by a limited knowledge of hyperon-hyperon and 
hyperon-nucleon three-body forces; see, however, a recent investigation 
of \citet{VidanaLPPB2011}, who estimated the effect of three-body 
forces in the hyperonic sector on the NS maximum mass. Assuming that 
hyperon interactions are weaker than the pure nucleonic interactions, 
the resulting non-rotating NS maximum mass is located in the $1.27 - 1.6\ M_\odot$ range. 
The problematic feature of low $M_{\rm max}$ is usually remedied by 
providing stronger repulsion between the hyperons and/or letting them appear
only at very high densities. Below, we list a selection of EOSs with hyperons,
derived from theories that successfully deal with the problem of a massive NS: 
the DS08 EOS \citep{DexheimerS2008} uses an effective hadronic SU(3) chiral model
including the baryon octet and fourth-order self-interaction terms of the
$\omega$, $\rho$, and $\phi$ vector mesons; the GM1Z0 EOS \citep{WeissenbornCS2012}
is based on a relativistic mean field model which allows the study of the departure
from a vector meson-hyperon couplings stemming from the SU(6) quark model to a
more general SU(3) prescription; the GM1 model with the ratio between the meson
octet and the singlet coupling constant $z=g_8/g_1=0$ (all the baryon-meson
couplings are equal) was used, which makes it the stiffest in our sample. The BM165
EOS \citep{BednarekHZBM2011} was also derived from a relativistic mean field
model, with a non-linear Lagrangian that includes quatric terms in the meson
fields, and two additional hidden-strangeness mesons $\sigma^*$ and $\phi$ that
couple to hyperons only. The TM1C EOS \citep{GusakovH2012} employs a prescription
similar to the BM165 EOS for purely nucleonic matter, and also introduces scalar
$\sigma^*$ and vector $\phi$ mesons in the description of hyperon interaction,
with an additional $\Lambda-\Lambda$ hyperon attraction. The SU(6) symmetry
breaking applied in the model amounts to $z=0.2$. 
%-----------------------------------------------------------------------------
%-----------------------------------------------------------------------------
\begin{table}[t]
\begin{center}
\caption{Selected parameters of static configurations (mass expressed 
in $M_\odot$, radius $R$ in km, the moment of inertia $I$ 
in $10^{45}\ {\rm g\cdot cm^2}$, and the frequency 
$f^{\rm s}=(1/2\pi)\sqrt{GM^{\rm s}_{\rm max}/R^3(M^{\rm s}_{\rm
max})}$ in Hz; 
$R_{1.4}$ and $I_{1.4}$ are computed for $M=1.4\ M_\odot$).}
\begin{tabular}[t]{ccccccc}
\hline
EOS & $M^{\rm s}_{\rm max}$ & $R(M^{\rm s}_{\rm max})$& $I^{\rm s}_{\rm max}$ 
& $R_{1.4}$ & $I_{1.4}$ & $f^{\rm s}$\\
\hline \hline
APR& 2.19 & 9.93 & 2.24 & 11.34 & 1.31 & 2740.47 \\ \hline 
BSK20& 2.17 & 10.17 & 2.00 & 11.75 & 1.39 & 2630.50 \\ \hline
DH& 2.05 & 9.99 & 2.27 & 11.73 & 1.37 & 2630.63 \\ \hline\hline
BM165& 2.03 & 10.68 & 2.24 & 13.59 & 1.74 & 2367.82 \\ \hline
DS08& 2.05 & 12.02 & 2.56 & 13.91 & 1.81 & 1993.42 \\ \hline
GM1Z0& 2.29 & 12.00 & 3.09 & 13.89 & 1.84 & 2108.51 \\ \hline
TM1C& 2.05& 12.51 & 2.72 & 14.51 & 1.91 & 1873.41 \\
\label{tab1}
\end{tabular}
\end{center}
\end{table}
%-----------------------------------------------------------------------------
%-----------------------------------------------------------------------------
\section{Results}
\label{sect:results}
%-----------------------------------------------------------------------------
%-----------------------------------------------------------------------------
The following subsections contain the results for constant spin frequency tracks
(sequences of configurations) of rotating stars. The frequency range spans 
an astrophysically-relevant range from $f=0$ Hz (static configurations, 
see Table~\ref{tab1} for their representative parameters) up to $f=1200$ Hz 
(i.e., much above the frequency of $716$ Hz of the most rapid pulsar to date,  
PSR J1748-2446ad of \citealt{Hessels2006}). Rigidly-rotating, stationary, 
and axisymmetric stellar configurations were obtained by means of the numerical library 
{\tt LORENE},\footnote{\tt http://www.lorene.obspm.fr} {\tt nrotstar} code, 
using the formulation of \citet{BonazzolaGSM1993}, with the accuracy checked by 
a 2D virial theorem \citep{BonazzolaG1994}. The sequences are
limited by the following conditions: from the low central density end, they
terminate at the so-called {\it mass-shedding} limit, when the orbital
frequency of a test particle at the star's equator equals the stellar spin
frequency $f$. The high central density limit is marked by the onset of the
axisymmetric perturbation instability described by the condition 
$\partial M/\partial \lambda|_J = 0$, where $M$ is the gravitational mass and 
$\lambda$ a suitable parametrization of the sequence of configurations 
(e.g., the central density; see \citealt{FriedmanIS1998} for details). 
We define the moment of inertia as $I=J/\Omega$, where $J$ is the total
stellar angular momentum and $\Omega$ is the angular frequency,
$\Omega = 2\pi f$ (for the definitions of $M$ and $J$, see 
\citealt{BonazzolaGSM1993}). For non-rotating configurations, $I$ is 
calculated using the slow-rotation approximation \citep{Hartle1967}.

Parameters of the approximate formul\ae\ presented below were obtained 
by the $\chi^2$ fitting, yielding a typical accuracy of the order of one per cent. 
%///////////////////////////////////////////////////////////////////////
\subsection{Gravitational mass}
\label{subsect:mass}
\begin{figure}[h]
\resizebox{\columnwidth}{!}
{\includegraphics{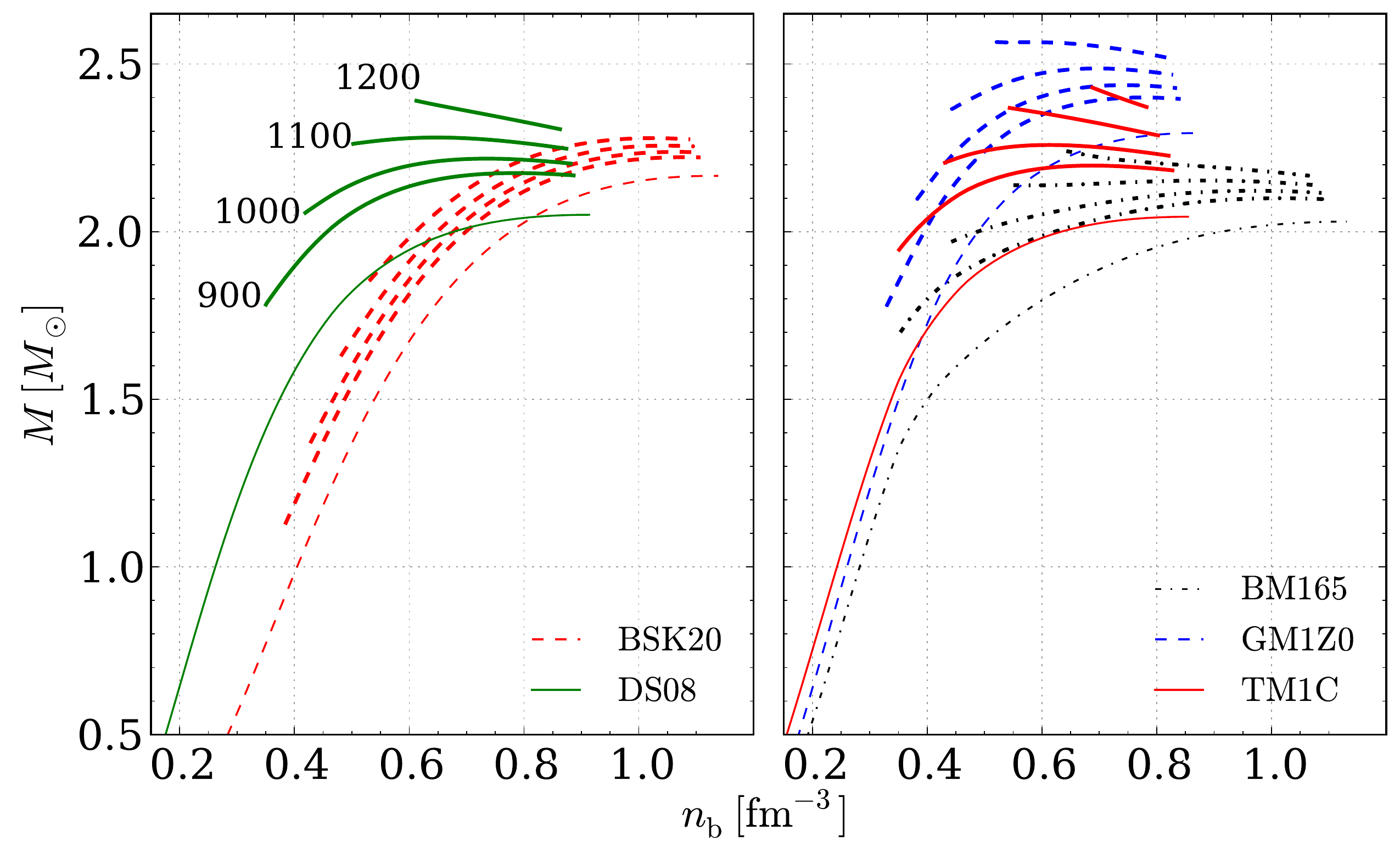}}
\caption{(Color online) Gravitational mass-central baryon density $M-n_{\rm b}$ relations 
for selected spin frequencies $f$, for stars composed of the BSK20 EOS 
(dashed red lines) and four hyperonic EOSs: DS08 (solid green), 
BM165 (dash-dotted black), GM1Z0 EOS (dashed blue) and TM1C (solid red lines). 
High- and low-density ends correspond to the axisymmetric instability and 
the mass-shedding limits, respectively.
From bottom to top for each EOS, $f$ equals 
0, 900, 1000, 1100 and 1200 Hz.}
\label{fig:m_f_nc}
\end{figure}

%///////////////////////////////////////////////////////////////////////
Figure~\ref{fig:m_f_nc} shows how the mass-central baryon density $M(n_{\rm b})$ 
relation changes with $f$ for the selected EOSs. A strong reduction of the available 
mass and the central density range for a large $f$ is characteristic of hyperonic EOSs; 
moreover, configurations near the mass-shedding limit (left sides of curves) may be 
more massive than those near the axisymmetric perturbation 
instability limit (right sides). 
%-----------------------------------------------------------------------
%-----------------------------------------------------------------------
%-----------------------------------------------------------------------
\begin{figure}[h]
\resizebox{\columnwidth}{!}
{\includegraphics{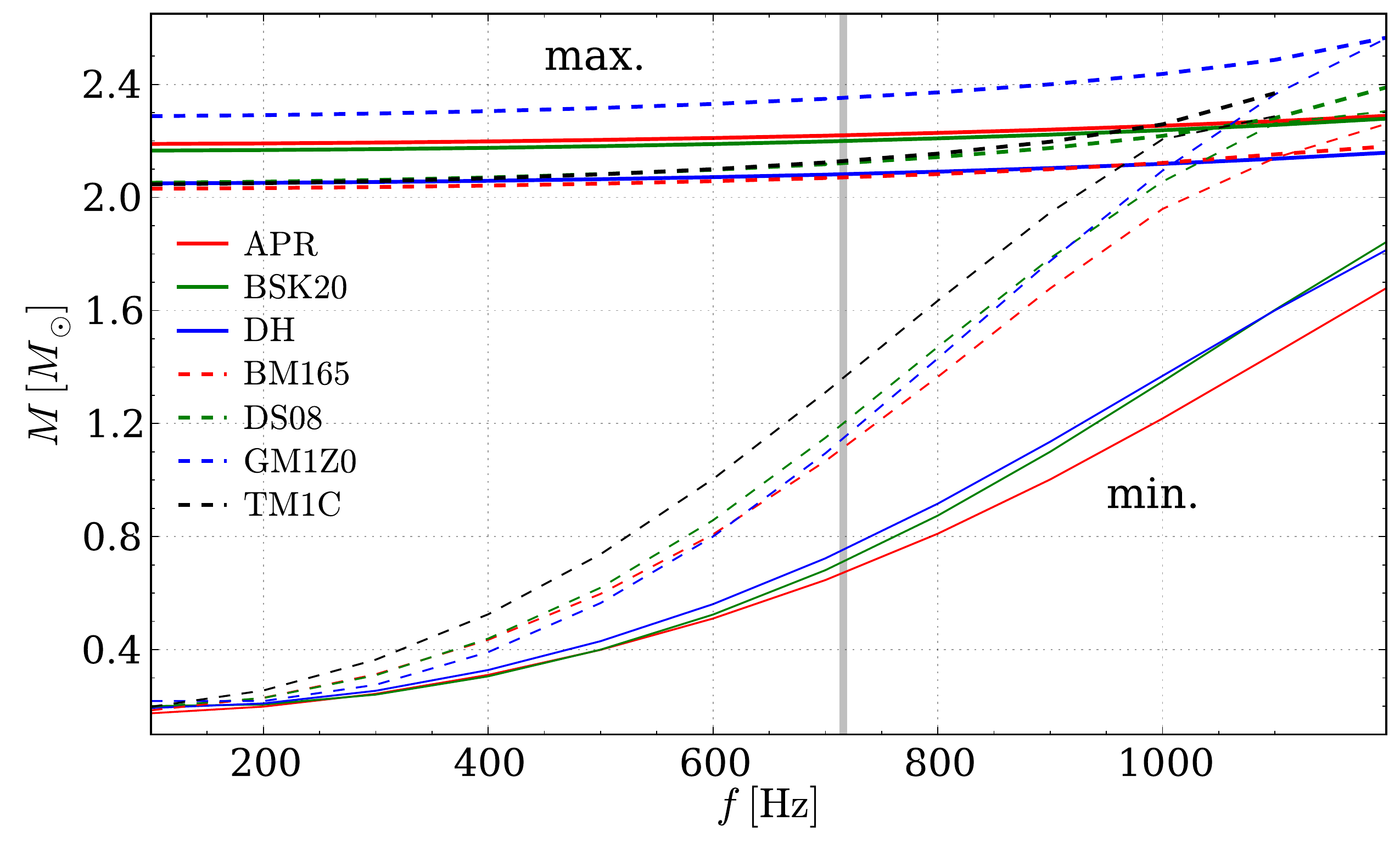}}
\caption{(Color online) Range of allowable gravitational masses as a function 
of spin frequency $f$. The vertical line indicates $716$
Hz frequency (spin frequency of the most-rapidly rotating NS to date).}
\label{fig:mminmax_f}
\end{figure}
%-----------------------------------------------------------------------
%-----------------------------------------------------------------------
\begin{figure}[h]
\resizebox{\columnwidth}{!}
{\includegraphics{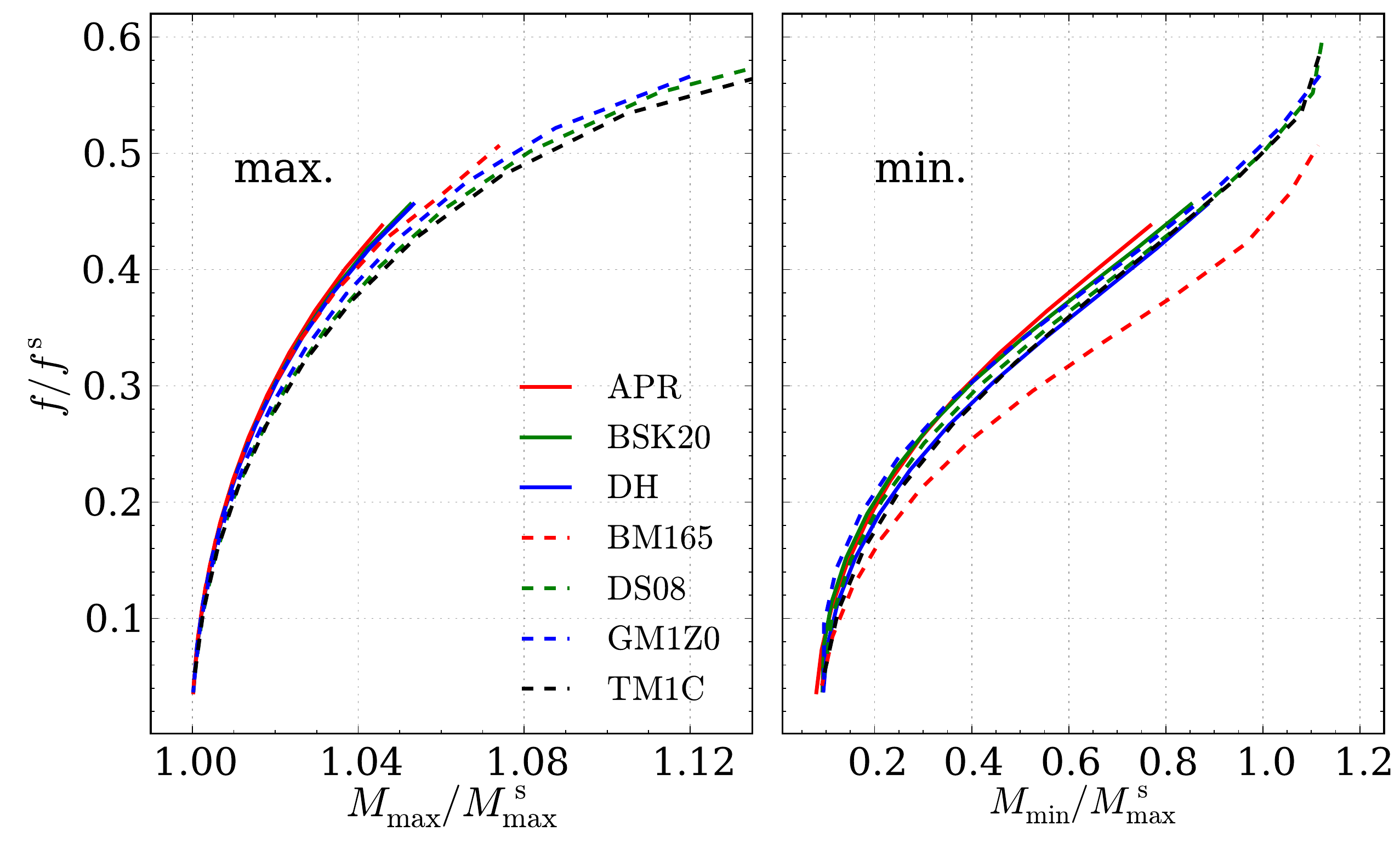}}
\caption{(Color online) Maximum mass
$M_{\rm max}$ (left panel) and minimum mass $M_{\rm min}$ (right panel) 
of rotating NSs as a function of the spin frequency $f$. 
Values are scaled by the parameters of non-rotating configurations.}
\label{fig:mimaf_scal}
\end{figure}
%-----------------------------------------------------------------------
Figure~\ref{fig:mminmax_f} shows the range of available masses as a function 
of spin frequency $f$; we note
that the mass of a hyperonic EOS star is already
strongly constrained by the existence of $716$ Hz pulsar \citep{Hessels2006}.
Detection of a $f\simeq 800$ Hz pulsar would mean that such a stiff EOS
cannot yield a star with $M<1.4\ M_\odot$. (In the
case of the TM1C EOS this minimum mass is already approached for 716 Hz.) 
The growth of the maximum mass $M_{\rm max}$, normalized by the maximum mass 
of the static configuration $M^{\rm s}_{\rm max}$ can be approximated 
by 
\begin{equation}
{M_{\rm max}}/{M^{\rm s}_{\rm max}} = 
0.49\left({f}/{f^{\rm s}}\right)^{8/3} + 1, 
\end{equation}
where $f^{\rm s} = (1/2\pi)\sqrt{GM^{\rm s}_{\rm max}/R^3(M^{\rm s}_{\rm
max})}$ is the frequency scaling factor\footnote{$f^{\rm s}$ equals the 
orbital frequency of a test particle orbiting spherical mass $M^{\rm s}_{\rm max}$ 
at a distance $R(M^{\rm s}_{\rm max})$.} formed of a static star maximum-mass 
parameters, gravitational mass $M^{\rm s}_{\rm max}$ and
corresponding radius $R(M^{\rm s}_{\rm max})$. In the case of minimum mass
$M_{\rm min}$, one can obtain a similar (although somewhat cruder)
approximation to the one for the $M_{\rm max}$,  
\begin{equation}
{M_{\rm min}}/{M^{\rm s}_{\rm max}} = 
3.57\left({f}/{f^{\rm s}}\right)^{2} + 0.1.
\end{equation}
The above relation underestimates the ${M_{\rm min}}/{M^{\rm s}_{\rm max}}$ 
for BM165 EOS stars for a given frequency, because 
the maximum-mass radius $R(M_{\rm max})$ 
is smaller than for other hyperonic EOSs, resulting in a larger 
$f^{\rm s}$.

%///////////////////////////////////////////////////////////////////////
\subsection{Surface redshifts}
\begin{figure}[h]
\resizebox{\columnwidth}{!}
{\includegraphics{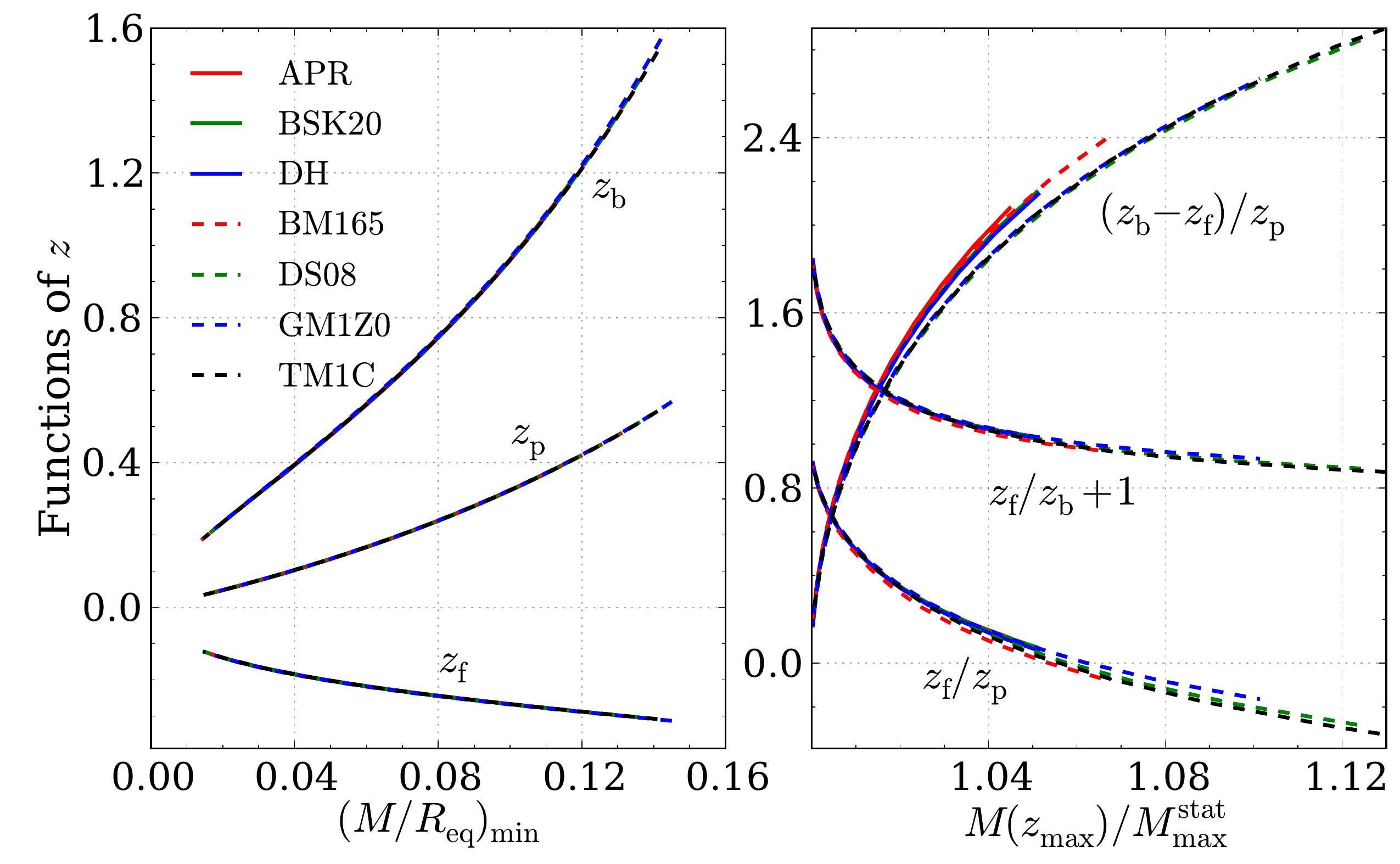}}
\caption{(Color online) Surface redshifts for rotating 
configurations. Left panel: Equatorial backward (top curve) and 
forward (bottom curve) redshifts $z_{\rm b}$ and $z_{\rm f}$, and 
the polar redshift $z_{\rm p}$ for configurations with the minimal 
mass, as functions of the compactness $(M/R_{\rm eq})_{\rm min}$.
Right panel: Functions of the three redshifts for rotating, 
maximally-compact configurations.}
\label{fig:z_f}
\end{figure}

%///////////////////////////////////////////////////////////////////////
Surface redshift may provide important information about the spacetime 
in the vicinity of the NS. In the case of a rotating star, one considers the 
redshift $z_{\rm p}$ of photons coming from the pole, as well as two equatorial 
redshifts for photons, emitted tangentially in and opposite the direction of rotation 
(forward $z_{\rm f}$ and backward $z_{\rm b}$ redshift, respectively; 
for definitions, see e.g., \citealt{Gourgoulhon2010}, Sect. 4.6). 
The left panel in Fig.~\ref{fig:z_f} shows the relations for $z$ for the minimum 
compactness $(M/R_{\rm eq})_{\rm min}$, i.e., configurations at the mass-shedding 
limit, that bear some resemblance to the configurations 
during the photospheric radius expansion 
burst (\citealt{SteinerLB2010} and references therein); $R_{\rm eq}$ denotes 
the equatorial radius. These relations may be regarded as 
{\it upper limits} on the compactness parameter for a measured $z$. 
We approximate $z_{\rm p}$ and $z_{\rm b}$ by $a\left(M/R\right)^{3/2} + b$, 
where $a=9.62$ and $b=0.023$ for $z_{\rm p}$, and $a=25.11$ and $b=0.18$ for $z_{\rm b}$; 
$z_{\rm f}$ is approximated by $-0.68\left(M/R\right)^{0.4}$. 

The right panel shows redshift functions for the {\it maximum-compactness} 
$(M/R_{\rm eq})_{\rm max}$ configuration along the line corresponding 
to the axisymmetric perturbation instability limit. For spin frequencies 
under consideration, it may be also treated as an approximate 
{\it lower limit} on the stellar mass.   
%-----------------------------------------------------------------------
%-----------------------------------------------------------------------
We propose parametrizing these relations as follows: 
for $\zeta = (z_{\rm b}-z_{\rm f})/z_{\rm p}$, 
$M/M^{\rm s}_{\rm max} \simeq 0.008\zeta^{2.5} + 1$; for  
$\zeta = 1-z_{\rm f}/z_{\rm p}$, $M/M^{\rm s}_{\rm max}$ is approximated by 
$0.06\zeta^{2.5} + 1$. In the case of the third function, we chose 
$\zeta = 1 - z_{\rm f}/z_{\rm b}$, and $M/M^{\rm s}_{\rm max}\simeq 
0.04\zeta^6 + 0.02\zeta^3 + 1$.  
%-----------------------------------------------------------------------
\begin{figure}[h]
\resizebox{\columnwidth}{!}
{\includegraphics{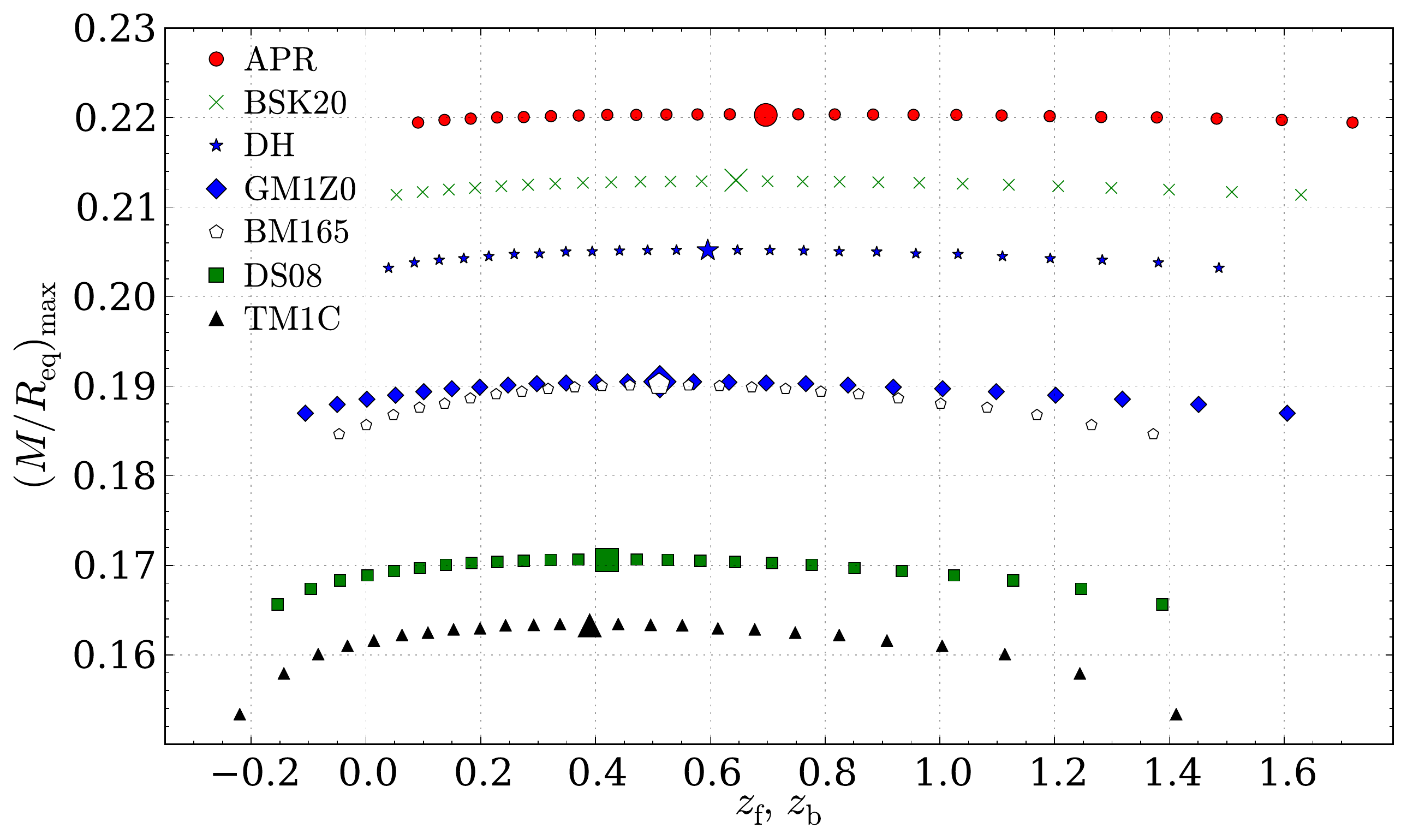}}
\caption{(Color online) Maximal compactness vs equatorial redshifts
$z_{\rm f}$ and $z_{\rm b}$ for different rotation rates; for 
$z_{\rm f}$, $f$ increases to the left of the bigger symbol that 
denotes the static configuration; for  $z_{\rm b}$, $f$ 
increases to the right. Points correspond to frequencies from 0 to 1200 Hz.}
\label{fig:z_comp_f}
\end{figure}
%-----------------------------------------------------------------------
The value of $z_{\rm b}-z_{\rm f}$ quantifies the redshift effect on the 
maximal line broadening coming from the equator of a rotating star. 
Figure~\ref{fig:z_comp_f} presents the maximal compactness for such stars 
against the two equatorial redshifts. The $(M/R_{\rm eq})_{\rm max}$ 
stays remarkably constant with rotation, and could be used 
to distinguish between the two sets of models.  
%///////////////////////////////////////////////////////////////////////
\subsection{Moment of inertia}
\begin{figure}[h]
\resizebox{\columnwidth}{!}
{\includegraphics{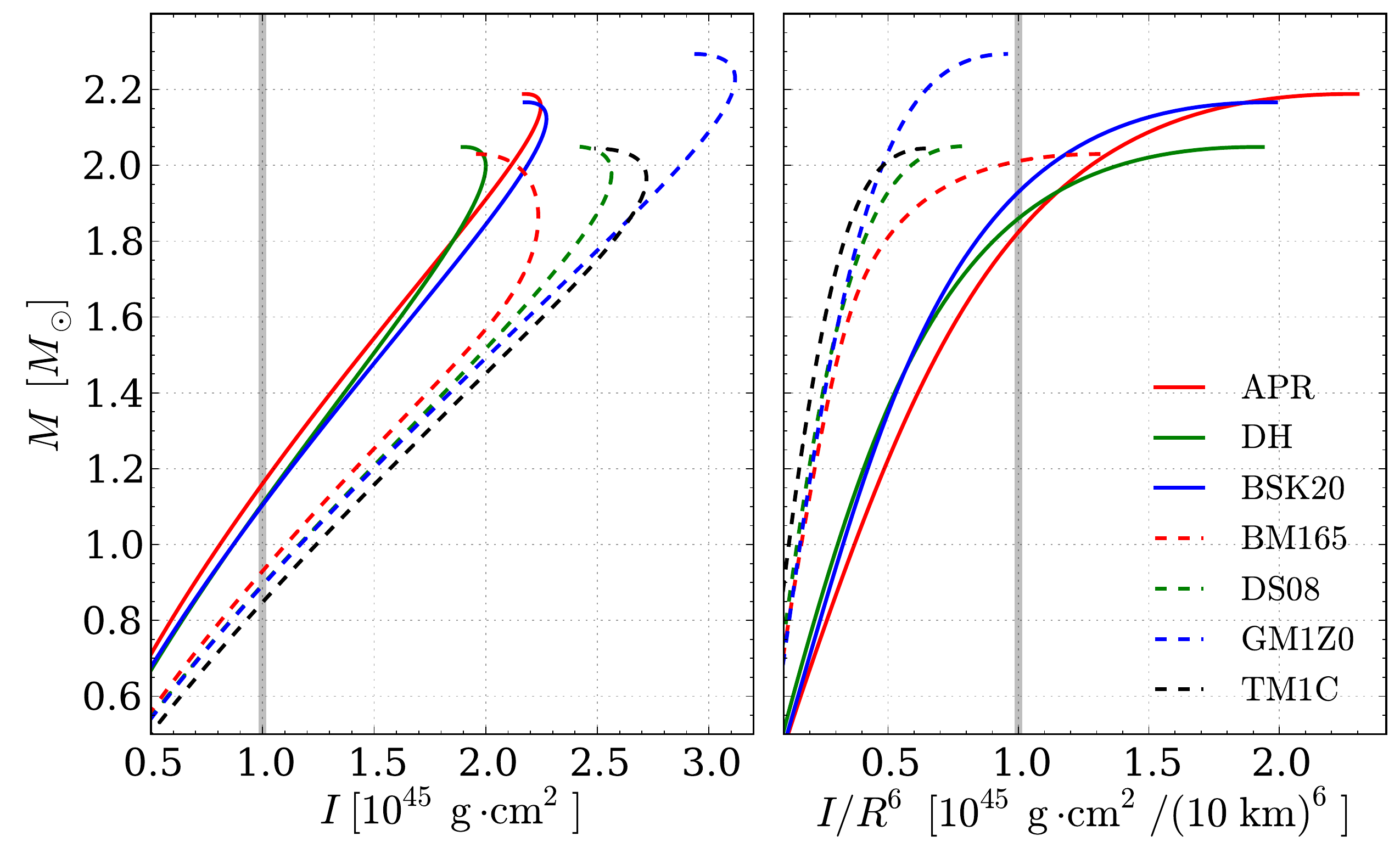}}
\caption{(Color online) Left panel: moment of inertia $I$ as a function
of gravitational mass $M$ for static solutions. Right panel: moment of
inertia $I$ divided by $R^6$ (as used in the estimation of the minimum 
surface magnetic field of a pulsar) for static stars. 
Vertical lines correspond to a fiducial model of 
$I=10^{45}\ {\rm g\cdot cm^2}$ and 10 km.} 
\label{fig:ivsm}
\end{figure}

\begin{figure}[h]
\resizebox{\columnwidth}{!}
{\includegraphics{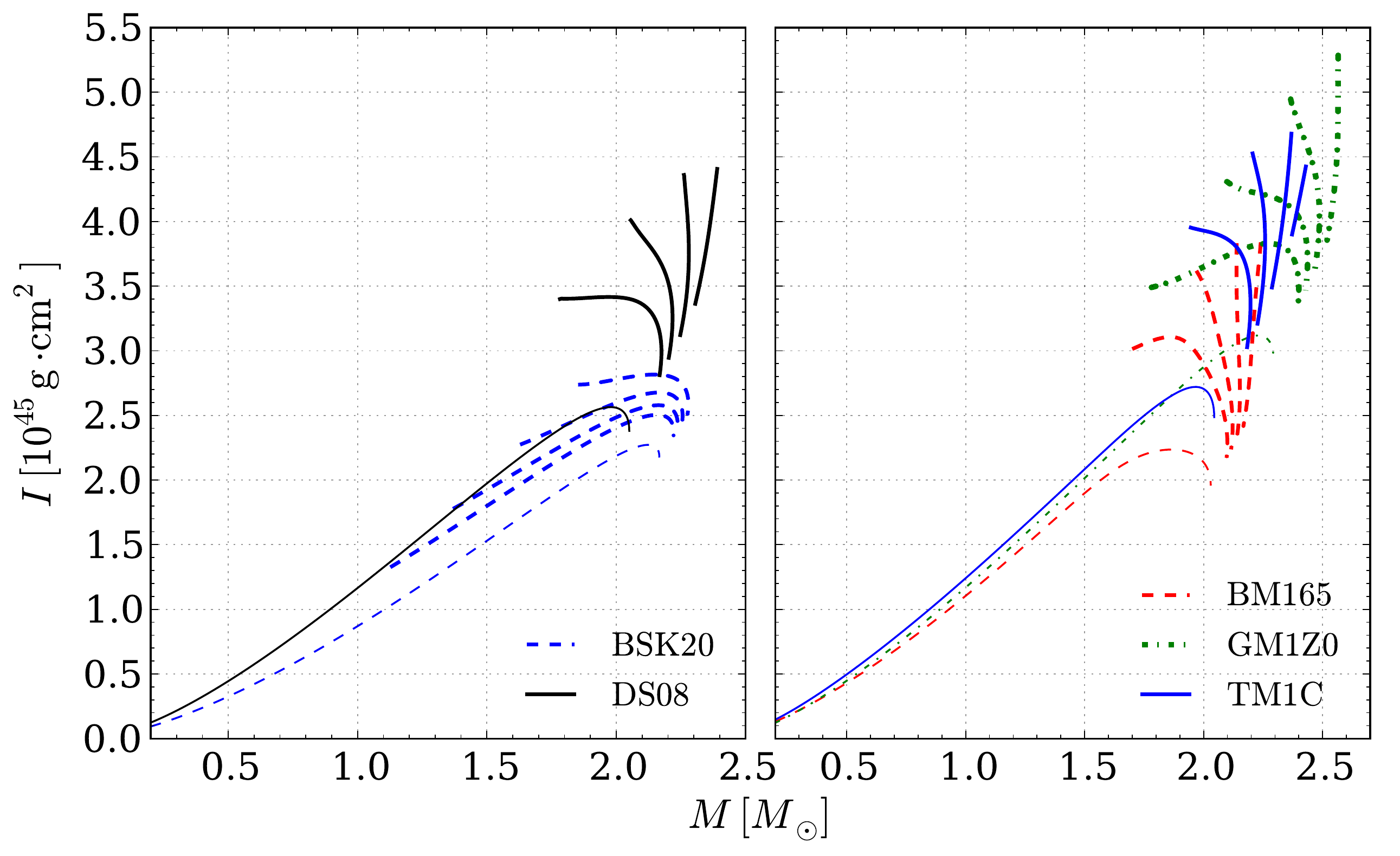}}
\resizebox{\columnwidth}{!}
{\includegraphics{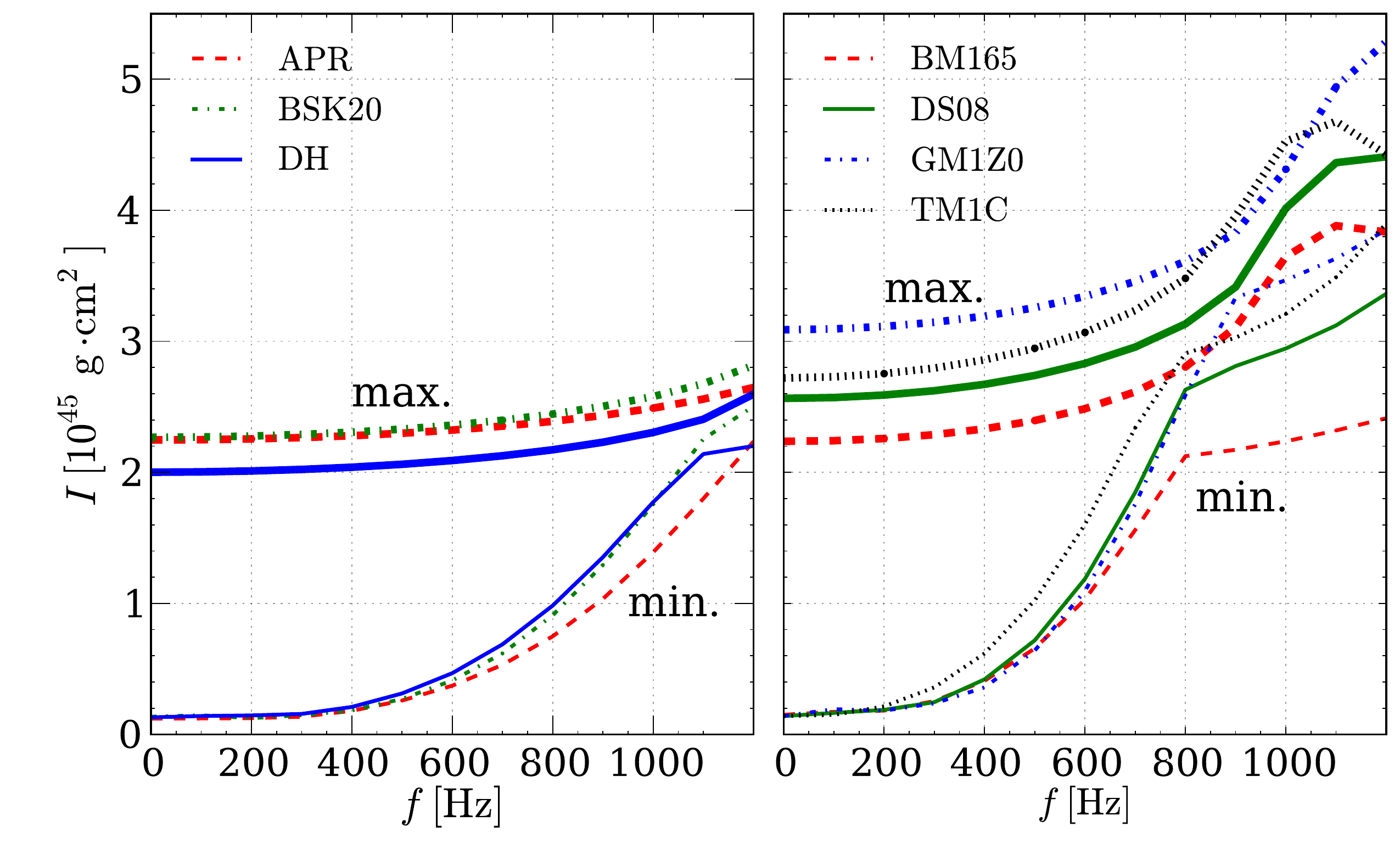}}
\caption{(Color online) Top panels: The behavior of $I(M)$ 
for rotating stars (from left to right $f$ equals 0, 900, 1000, 1100 
and 1200 Hz) for hyperonic EOSs and DH EOS for comparison. Bottom panels: 
Allowable range of the moment of inertia for purely nucleonic 
stars (left) and hyperonic stars (right). Thin (thick) 
lines correspond to the minimum (maximum) of $I$ that a stable configuration 
can attain at a given frequency.}
\label{fig:imin_imax_f}
\end{figure}

%///////////////////////////////////////////////////////////////////////
A potentially interesting NS parameter is the  
moment of inertia $I$ (in principle measurable using the spin-orbit 
coupling, \citealt{LattimerS2005}). 
Figure~\ref{fig:ivsm} shows the behavior of the gravitational mass $M$ 
as a function of the moment of inertia $I$ for static configurations. 
The curves in the left panel may be approximated by a straight line, 
$I(M) = a_{\rm I}M + b_{\rm I}$. For nucleonic EOSs, $a_{\rm I} = 1.22$ 
and $b_{\rm I} = -0.34$, for hyperonic EOSs $a_{\rm I} = 1.51$.  
In the right panel we plot the value of $I/R^6$ as a function of $M$; 
this value enters the estimation of the minimum magnetic dipole field 
at the pulsar surface, 
$B > \left(I/R^6\right)^{1/2}\left(3c^3\dot{P}P/8\pi^2\right)^{1/2}$, 
where $P=2\pi/f$ denotes the spin period\footnote{Although 
$I$ grows with rotation, the value of $I/R^6$ decreases
because of the high power of $R$ in the denominator. The static results
are therefore upper limits for the value of $I/R^6$ for a given $M$.}. 
Figure~\ref{fig:imin_imax_f} presents how the available range of the moments 
of inertia changes with rotation. For the frequencies $f \ge 800$ Hz  
the moment of inertia tend to be already quite constrained. The minimum 
value of $I$ for sub-millisecond rotation coincides with the high-density 
end of the track, while the mass-shedding limit is characterized by a 
large $I$ (because of a strong equatorial radius dependence, the effect is much 
more pronounced than for the gravitational mass, Fig.~\ref{fig:mminmax_f}).  
%-----------------------------------------------------------------------
%-----------------------------------------------------------------------
%-----------------------------------------------------------------------
%-----------------------------------------------------------------------
%-----------------------------------------------------------------------
\begin{figure}[h]
\resizebox{\columnwidth}{!}
{\includegraphics{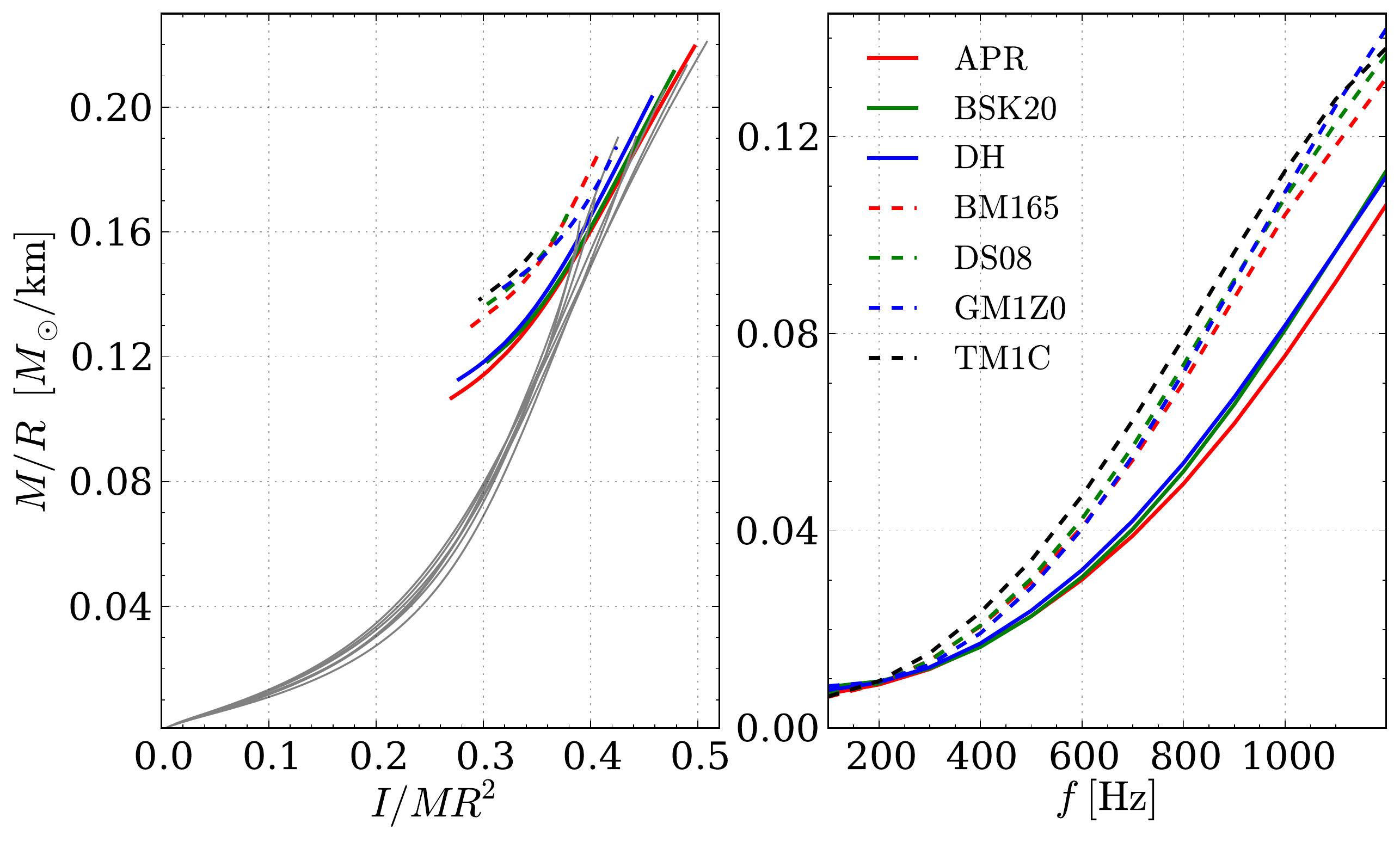}}
\caption{(Color online) Left panel: relation between $I/MR^2$ and the
compactness parameter $M/R$ for static configurations (grey) and those rotating at
1200 Hz, in which case $R \equiv R_{\rm eq}$. Right panel: Minimal $M/R_{\rm
eq}$ ratio (value corresponding to the mass-shedding configuration) as a
function of rotational frequency.}
\label{fig:ivsmr}
\end{figure}
%-----------------------------------------------------------------------
As many softer EOSs are now ruled out by observations, we reproduce the 
formula suggested by \citet{LattimerP2001} in the form of \citet{BejgerH2002}, 
with slightly different coefficients in the static case.  
For an astrophysically relevant range of masses $M>0.5\ M_\odot$    
we have 
\begin{equation}
I/MR^2 = a_0\left({M}/{R}\right) + b_0, 
\label{eq:ivsmr}
\end{equation}
with $a_0 = 1.40$ and $b_0 = 0.19$ (left panel in Fig.~\ref{fig:ivsmr}). 
This approximation is generally valid for the static results 
in our sample. A more universal formula, which takes rotation into account, 
will depend on the spin frequency $f$; the compactness $M/R$ 
must be re-defined as the  
gravitational mass $M$ to the circumferential radius $R_{\rm eq}$ ratio.  
For EOSs without hyperons, we propose the following approximation: 
\begin{equation}
a(f) = 0.19 f_1^{4} + a_0~~~~{\rm and}~~~~
b(f) = -0.042 f_1^{4} + b_0,  
\end{equation}
where $f_1 = f/({\rm 1\ kHz})$. Since the rotation influences hyperonic 
stars in a different way, we produce an analogous approximation for a subset 
of hyperonic EOSs only. In the static case we obtain $a_{0h} = 1.14$ 
and $b_{0h} = 0.22$, and for rotating configurations
\begin{equation}
a(f) = 0.51 f_1^{4} + a_{0h}~~~~{\rm and}~~~~
b(f) = -0.099 f_1^{4} + b_{0h}. 
\end{equation}
In order to estimate the moment of inertia, one needs to 
know the {\it minimum} value of the compactness parameter 
$M/R_{\rm eq}$ for a given spin frequency (right panel in 
Fig.~\ref{fig:ivsmr}). It can be described by 
\begin{equation}
\left({M}/{R_{\rm eq}}\right)_{\rm min} = 
a_{\rm c}f_1^2 + b_{\rm c}, 
\end{equation}
where $b_{\rm c}=0.005$, and $a_{\rm c}$ depends on the type
of the EOS. For EOSs without hyperons $a_{\rm c}=0.074$, otherwise 
$a_{\rm c}=0.099$. Thanks to their stiffness, hyperonic EOSs produce larger 
radii near the mass-shedding limit, but the minimum mass for a given $f$ 
is also higher in comparison to nucleonic EOSs 
(see Fig.~\ref{fig:mminmax_f}), which results in higher $(M/R_{\rm eq})_{\rm min}$.  
%-----------------------------------------------------------------------
%///////////////////////////////////////////////////////////////////////
\subsection{Spin-up by disk accretion}
\label{subsect:acc}
%///////////////////////////////////////////////////////////////////////
A process in which the moment of inertia is relevant is 
the so-called {\it recycling} of pulsars to millisecond periods. 
To study the relation of spin-up to the EOS we use the model described 
in \citet{BejgerFHZ2011}, a magnetic torque in the form given by 
\citealt{KluzniakR2007}, updated to include the marginally-stable orbit  
and the magnetic field decay proportional to the amount of accreted 
mass. The evolution of the total stellar angular momentum in the 
process of accretion is 
\begin{equation}
{{\rm d}J}/{{\rm d}M_b} = l - l_{\rm mag} = l_{\rm tot}, 
\label{eq:djdmb}
\end{equation} 
where $l$ is the orbital angular momentum of a particle in the disk
per unit baryon mass, and $l_{\rm mag}$ is the contribution from the {\it
braking} magnetic torque, resulting from the magnetosphere--disk interaction. 

For comparison with the realistic EOSs, we employ a useful fiducial 
model:\footnote{In general, the changes of stellar parameters relate to each
other as ${\rm d}M = \Omega {\rm d}J + \mu{\rm d}M_b$, where $\mu = 1/u^t$ is the
chemical potential per unit mass, transfered onto the star by the accreting
particle, $u^t$ being the time component of the particle four-velocity. In the 
case of $B=0$, $\mu = e - \Omega l$, where $e$ is the specific energy 
of a particle infalling onto the star \citep{SibgatillinS2000,ZdunikHG2002}, the
gravitational mass growth of such a star is therefore 
${\rm d}M = e{\rm d}M_b$; we adopt this prescription for $B\neq 0$ too (we also correct 
a misprint in Eqs. 4 and 5 of \citealt{BejgerFHZ2011}: instead of $u^t$, it should 
read $1/u^t$).} 
I45, a star with a {\it constant} radius $R=10$ km, and a {\it constant}
moment of inertia $I=10^{45}\ {\rm g\cdot cm^2}$. 
%-----------------------------------------------------------------------
\begin{figure}[h]
\resizebox{\columnwidth}{!}
{\includegraphics{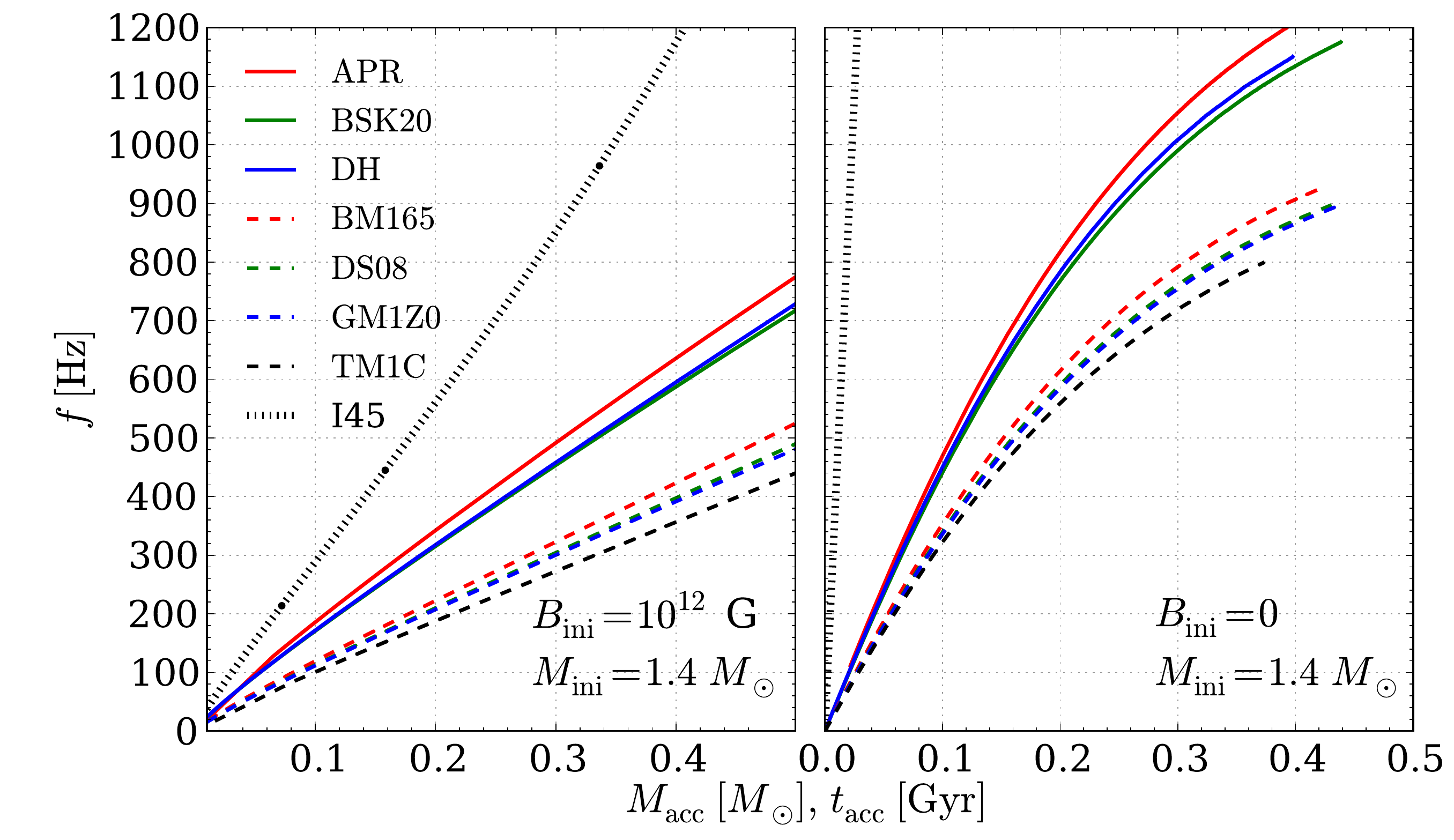}}
\caption{(Color online) Spin evolution of an accreting star 
as a function of the accreted mass (accretion rate is 
$\dot{M}=10^{-9}\ M_\odot/{\rm yr}$, the horizontal axes correspond to 
accretion time $t_{\rm acc}$ in Gyr) with initial mass $M_{\rm ini}=1.4\ M_\odot$. 
Left panel: accretion with the magnetic field decay and $B_{\rm ini}=10^{12}$ G. 
Right panel: accretion without $B=0$ (from the marginally-stable orbit).}
\label{fig:acc_nobb14}
\end{figure}
%-----------------------------------------------------------------------
%-----------------------------------------------------------------------
\begin{figure}[h]
\resizebox{\columnwidth}{!}
{\includegraphics{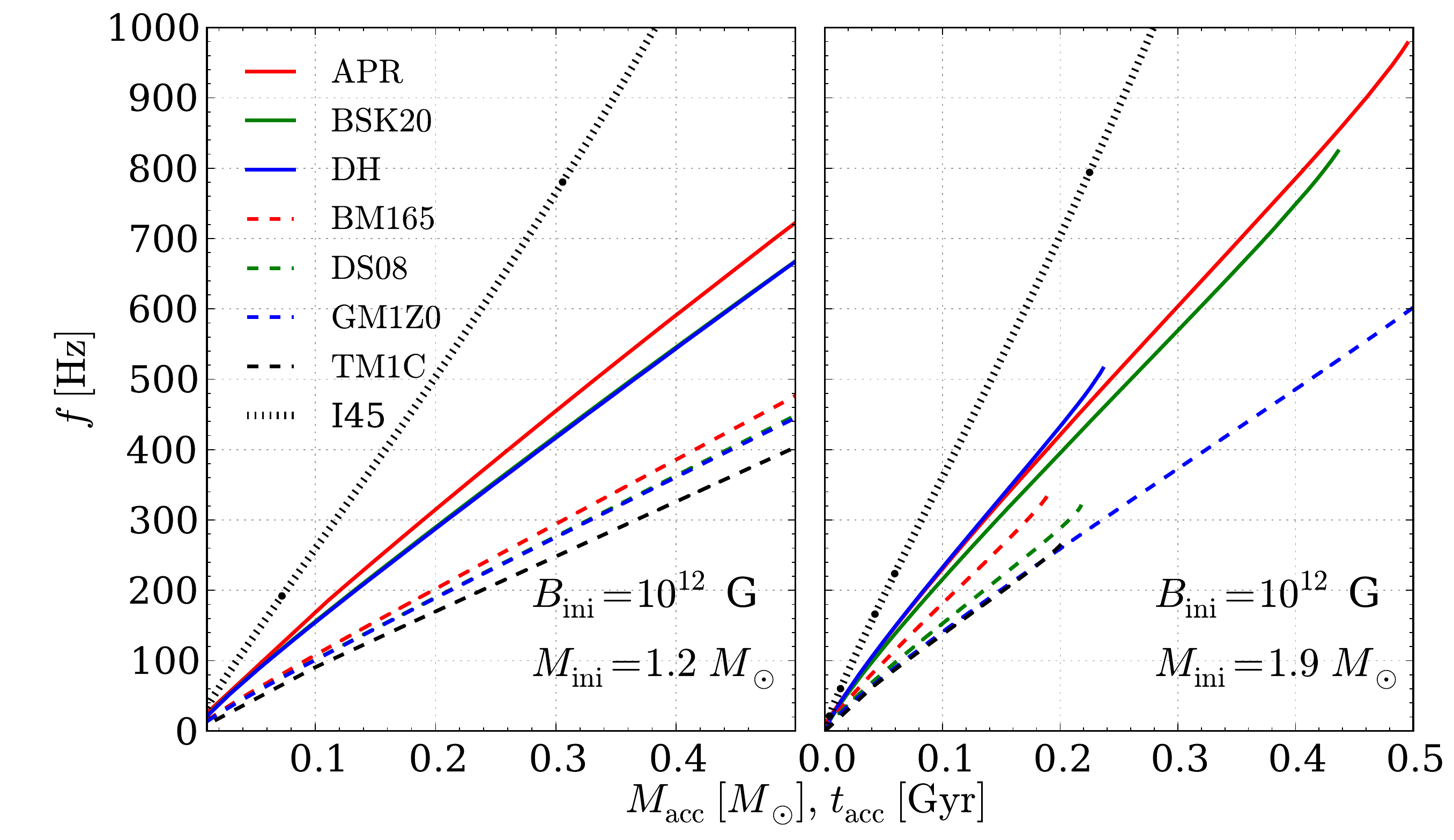}}
\caption{(Color online) As in the left panel of Fig.~\ref{fig:acc_nobb14}, 
but for $M_{\rm ini}=1.2\ M_\odot$ (left panel) and $M_{\rm ini}=1.9\ M_\odot$ 
(right panel).} 
\label{fig:acc_b12_19}
\end{figure}
%-----------------------------------------------------------------------
To compare the effect of the EOS on the recycling process, Fig.~\ref{fig:acc_nobb14} 
shows the evolution of spin frequency $f$ for a chosen accretion rate,  
$\dot{M}=10^{-9}\ M_\odot/{\rm yr}$ and initial magnetic field $B=10^{12}$ G 
(left panel). As anticipated, stars with larger moments of inertia gain $f$ 
less rapidly. With the above parameters, a hyperonic star has to accrete almost 
twice as much matter as a nucleonic EOS star to reach given $f$ (see also 
\citealt{BejgerFHZ2011}). For comparison, the 
right panel shows the importance of the magnetic 
torque. When it is neglected ($B=0$), the amount of mass needed to 
spin up the star to a given $f$ is much smaller (lines end at the 
mass-shedding limit). By rewriting Eq.~\ref{eq:djdmb} for $J=I\Omega$, 
\begin{equation}
{{\rm d}\Omega}/{{\rm d}M_b} = \left(l_{\rm tot} 
- \Omega{{\rm d}I}/{{\rm d}M_b}\right)/I, 
\label{eq:dodmb}
\end{equation}
one deduces that both large $I$ and its growth with $M$ decrease the 
spin-up rate (to be compared with the fiducial I45 configuration; see also 
the top-left panel in Fig.~\ref{fig:ivsm} where we compare the BSK20 and DS08 EOSs). 
We also note that, contrary to intuition, more massive stars are spinning up 
faster, as shown in Fig.~\ref{fig:acc_b12_19} for $M_{\rm ini}=1.2\ M_\odot$ 
and $M_{\rm ini}=1.9\ M_\odot$. This feature results from the 
fact that $l_{\rm tot}$ is larger for larger $M$. In the general case,  
the expression for $l_{\rm tot}(M)$ is rather complicated (see \citealt{BejgerFHZ2011} 
for details); however, for $B=0$ in the Schwarzschild case, 
$l\propto \sqrt{Mr_{\rm ms}}$, and $r_{\rm ms}$, the radius of the innermost 
stable circular orbit, depends linearly on $M$. This effect is independent 
of the change of $I$, as shown for the $I=const.$ tracks 
with different $M_{\rm ini}$. The change of ${{\rm d}I}/{{\rm d}M_b}$ for 
realistic EOSs, related to the $I(M)$ behavior near the $M_{\rm max}$, is visible 
as a slight change of slope in the right panel of Fig.~\ref{fig:acc_b12_19}, just 
before the accretion ends at the instability limit.  
%///////////////////////////////////////////////////////////////////////
\subsection{Comparison with stiff nucleonic EOS}
\label{subsect:stiffcomp}
\begin{figure}[h]
\resizebox{\columnwidth}{!}
{\includegraphics{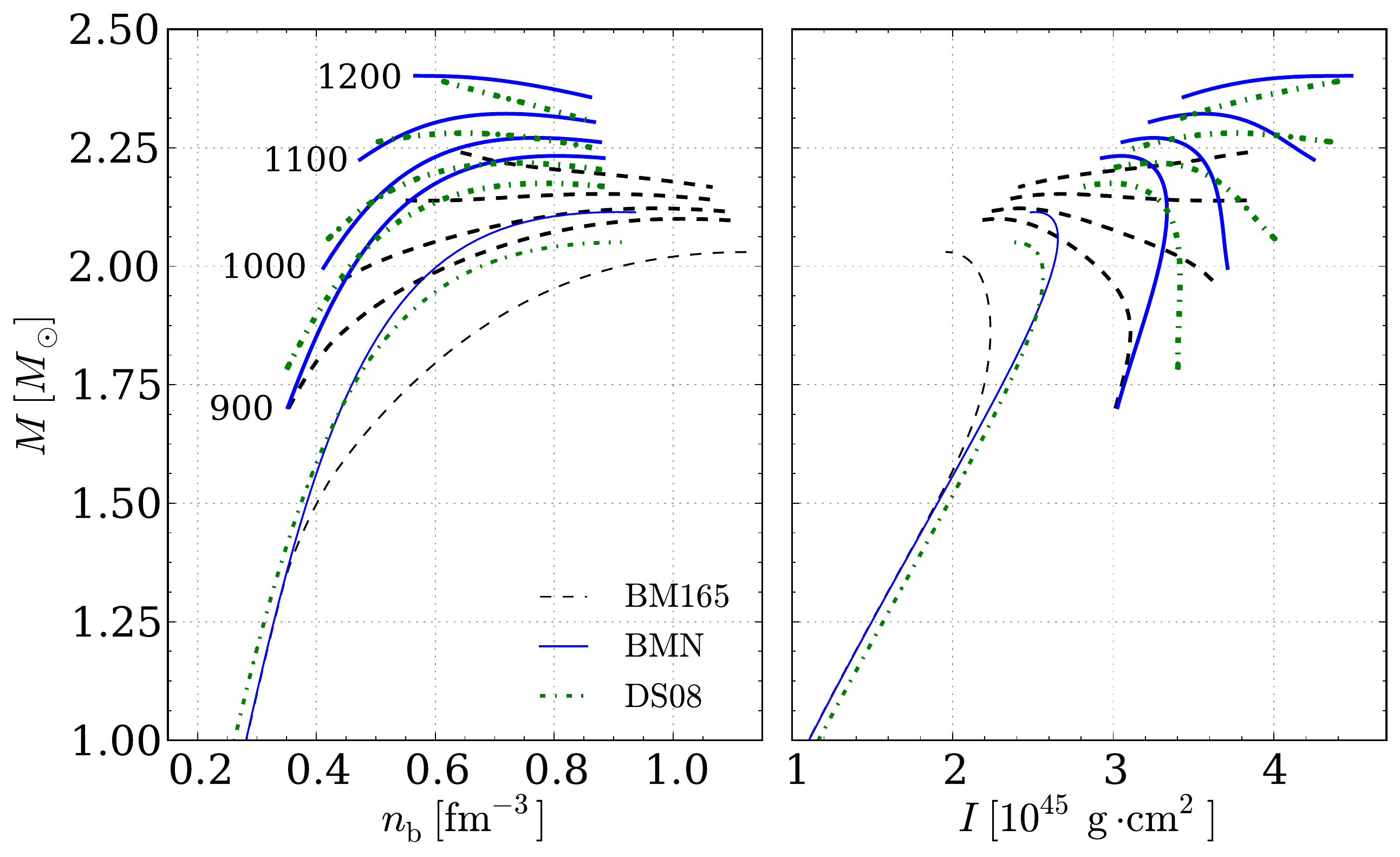}}
\caption{(Color online) Gravitational mass-central baryon density $M-n_{\rm b}$  
and the moment of inertia-mass $I(M)$ relations for BM165 (dashed black), 
BMN (solid blue), and DS08 EOS (dash-dotted green lines).}
\label{fig:bmcomp}
\end{figure}

%///////////////////////////////////////////////////////////////////////
To be consistent with the $2\ M_\odot$ NS mass measurement, the hyperonic
EOSs are necessarily stiff for densities lower than the hyperon appearance
threshold. This requirement is not essential for nucleonic EOSs, and is directly
reflected in higher compactnesses, smaller radii and smaller 
moments of inertia of associated stellar
models. The nucleonic EOS {\it may} however be stiff; to investigate the
imprint of hyperons, we construct a purely nucleonic BMN EOS by suppressing 
hyperons in the BM165 EOS (see \citealt{BednarekHZBM2011} for details, where
the BMN EOS is called EOS.N). Compared to the BM165 EOS, it yields a moderately 
higher non-rotating maximum mass, $M^{\rm s}_{\rm max}=2.11\ M_\odot$ and 
larger $R(M^{\rm s}_{\rm max})$, 11.96 km. The lack of hyperons results in 
available central density ranges that are shifted towards lower values 
for a given spin frequency, as well as higher moments of inertia 
(see Fig.~\ref{fig:bmcomp}, in which the left and right panels correspond to 
Figs.~\ref{fig:m_f_nc} and \ref{fig:imin_imax_f}).

Because of the considerable freedom in choosing EOS parameters, the discrimination
between a given hyperonic EOS and another stiff, but purely nucleonic EOS is
notably hindered, mostly because they give similar values of the stellar 
radius that affects potentially-measurable parameters.   
For relations presented in Figs.~\ref{fig:m_f_nc}-\ref{fig:acc_b12_19}, 
the BMN EOS results resemble those of the DS08 EOS (e.g., in 
Fig.~\ref{fig:z_comp_f} the central value of $(M/R_{\rm eq})_{\rm max}$ equals 0.177). 

One can thus, in principle, construct a nucleonic EOS that will result 
in stellar parameters similar to hyperonic models; the reverse statement is, 
at least up to the current state of the art, not true. 
%-----------------------------------------------------------------------
%-----------------------------------------------------------------------
%///////////////////////////////////////////////////////////////////////
\section{Summary and conclusions}
\label{sect:conc}
%///////////////////////////////////////////////////////////////////////
We have studied a set of recent EOSs that contain hyperons and yield 
static NSs consistent with a robust observational constraint, 
a $2 M_\odot$ NS mass measured by \citet{Demorest2010}. 
Stars constructed using hyperonic EOSs were compared with those using 
a minimalistic approach, i.e., containing nucleons only. 
Their global parameters, gravitational mass, compactness, surface redshift, 
and moments of inertia were calculated for a broad 
range of spin frequencies ($0-1200$ Hz). 
We have focused on the extreme values of these parameters and find 
that the minimum mass for hyperonic EOSs increases much 
faster with rotation than in the case of the representative set of  
nucleonic EOSs described in Sect.~\ref{sect:meth_eos}. 
For frequencies just slightly larger than 716 Hz the minimum mass 
approaches $1.4\ M_\odot$. In the case of the compactness parameter  
$M/R_{\rm eq}$, its minimal value is substantially larger than for nucleonic    
EOSs at a given $f$ (Fig.~\ref{fig:ivsmr}). We also note that the  
sub-millisecond rotation confines the mass into a narrow interval 
(a feature that is present for softer EOS stars for much higher 
frequencies, see \citealt{BejgerHZ2007}). However, as shown 
in Sect.~\ref{subsect:stiffcomp}, stellar models based on a sufficiently 
stiff nucleonic EOS may be confused with hyperonic EOS models; 
in view of the measurement of \citet{Demorest2010}, 
an observation of a compact NS (i.e., suggesting a softer nucleonic EOS) 
with a mass far below $2\ M_\odot$ would make the existence of hyperons 
at higher densities less plausible. 

A number of approximate formul\ae\ describing the whole set of EOSs is
provided. We approximate the relations between the $M_{\rm max}$, 
$M_{\rm min}$, $(M/R_{\rm eq})_{\rm min}$, and functions of surface redshift for
minimal and maximal compactness configurations for a range of studied
frequencies. We find that the maximal compactness configuration has 
$M/R_{\rm eq}$ that changes little with rotation.

Because of their stiffness, hyperonic stars have comparatively large moments of
inertia that increase with rotation to values significantly larger 
than the usual $10^{45}\ {\rm g\cdot cm^2}$. We extend the parametrization 
of \citealt{BejgerH2002} by supplying an approximate $I/(MR^2)$ vs $M/R$ 
relation for rotating stars. A large moment of inertia is also one of the 
parameters in the process of recycling a millisecond pulsar 
that hinders the spin-up. If hyperons exist in the interior of NSs (i.e., the 
nucleonic matter is sufficiently stiff), it may be one of the reasons high 
frequencies are not observed.
%///////////////////////////////////////////////////////////////////////
\begin{acknowledgements}
I thank N. Chamel, D. Chatterjee, V. Dexheimer, and M. E. Gusakov for 
providing their tabulated EOSs, and P. Haensel, J. L. Zdunik, and 
an anonymous referee for comments. 
This work was partially supported by the Polish MNiSW
research grants no. 2011/01/B/ST9/04838, N N203 512838 and completed using 
free and open software ({\tt LORENE}, {\tt gnuplot}, {\tt matplotlib}, 
{\tt numpy} and {\tt scipy}).
\end{acknowledgements}

%-----------------------------------------------------------------------


\begin{thebibliography}{}

\bibitem[Akmal et al.(1998)]{AkmalPR1998}
Akmal A., Pandharipande V. R., Ravenhall D. G., 1998,
Phys. Rev. C, 58, 1804

\bibitem[Bednarek et al.(2011)]{BednarekHZBM2011}
Bednarek, I. et al., 2011, \aap, 543, A157

\bibitem[Bejger \& Haensel(2002)]{BejgerH2002} 
Bejger, M., \& Haensel, P.\ 2002, \aap, 396, 917 

\bibitem[Bejger et al.(2007)]{BejgerHZ2007}
Bejger, M., Haensel, P., \& Zdunik, J.~L.\ 2007, \aap, 464, L49

\bibitem[Bejger et al.(2010)]{BejgerZH2010}
Bejger M., Zdunik J. L., Haensel P., 2010, \aap, 520, A16

\bibitem[Bejger et al.(2011a)]{BejgerFHZ2011}
Bejger M., Fortin M., Haensel P., Zdunik J. L., 2011 \aap, 536, A87

\bibitem[Bejger et al.(2011b)]{BejgerHZF2011}
Bejger M., Haensel P., Zdunik J. L., Fortin M., 2011 \aap, 536, A92

\bibitem[Bonazzola et al.(1993)]{BonazzolaGSM1993}
Bonazzola, S., Gourgoulhon, E., Salgado, M., \& Marck, J.~A.\ 1993, \aap, 278, 421

\bibitem[Bonazzola \& Gourgoulhon(1994)]{BonazzolaG1994}
Bonazzola, S., \& Gourgoulhon, E.\ 1994, CQG, 11, 1775

\bibitem[Burgio et al.(2011)]{BurgioSL2011}
Burgio, G.F., Schulze, H.-J., Li, A., 2011, Phys Rev C, 83, 025804

\bibitem[Demorest et al.(2010)]{Demorest2010}
Demorest P. B., et al., 2010, Nature, 467, 1081

\bibitem[Dexheimer \& Schramm(2008)]{DexheimerS2008}
Dexheimer, V., \& Schramm, S.\ 2008, \apj, 683, 943

\bibitem[Douchin \& Haensel(2001)]{DouchinH2001}
Douchin F. \& Haensel P., 2001, \aap, 380, 151

\bibitem[Friedman et al.(1988)]{FriedmanIS1998}
Friedman, J.L., Ipser, J.R, Sorkin, R.D., 1988, ApJ, 325, 722

\bibitem[Gourgoulhon(2010)]{Gourgoulhon2010}
Gourgoulhon, E.\ 2010, arXiv:1003.5015 

\bibitem[Goriely et al.(2010)]{GorielyCP2010}
Goriely, S., Chamel, N., \& Pearson, J. M. 2010, Phys.Rev.C, 82, 035804

\bibitem[Gusakov et al.(2012)]{GusakovH2012}
Gusakov, M. E., Haensel, P., Kantor, E. M., {\it private communication}

\bibitem[Haensel et al.(2007)]{HaenselPY2007}
Haensel, P., Potekhin, A.~Y., \& Yakovlev, D.~G.,  
''Neutron Stars 1: Equation of State and Structure'', New York: Springer, 2007
%ASSL, 326, Astrophysics and Space Science Library, 326

\bibitem[Hartle(1967)]{Hartle1967}
Hartle, J.~B.\ 1967, \apj, 150, 1005

\bibitem[Hessels et al.(2006)]{Hessels2006} 
Hessels, J.~W.~T., et al.\ 2006, Science, 311, 1901 

\bibitem[Klu{\'z}niak \& Rappaport(2007)]{KluzniakR2007}
Klu{\'z}niak W. \& Rappaport S., 2007, ApJ, 671, 1990

\bibitem[Lattimer \& Prakash(2001)]{LattimerP2001}
Lattimer J.M. \& Prakash M., 2001, ApJ, 550, 426

\bibitem[Lattimer \& Schutz(2005)]{LattimerS2005}
Lattimer, J.~M., \& Schutz, B.~F.\ 2005, \apj, 629, 979 

\bibitem[Schulze et al.(2006)]{SchulzePRV2006}
Schulze, H.-J., Polls, A., Ramos, A., \& Vida{\~n}a, I.\ 2006, \prc, 73, 058801

\bibitem[Sibgatullin \& Sunyaev(2000)]{SibgatillinS2000} 
Sibgatullin, N.~R., \& Sunyaev, R.~A.\ 2000, Astronomy Letters, 26, 772

\bibitem[Steiner et al.(2010)]{SteinerLB2010} 
Steiner, A.~W., Lattimer, J.~M., \& Brown, E.~F.\ 2010, \apj, 722, 33

\bibitem[Vida{\~n}a et al.(2011)]{VidanaLPPB2011} 
Vida{\~n}a, I., et al. %Logoteta, D., Provid{\^e}ncia, C., Polls, A., \& Bombaci, I.
\ 2011, EPL (Europhysics Letters), 94, 11002

\bibitem[Weissenborn et al.(2012)]{WeissenbornCS2012}
Weissenborn, S., Chatterjee, D., \& Schaffner-Bielich, J.\ 2012, \prc, 85, 065802

\bibitem[Zdunik et al.(2002)]{ZdunikHG2002}
Zdunik, J.~L., Haensel, P., \& Gourgoulhon, E.\ 2002, \aap, 381, 933

\end{thebibliography}
\end{document}